\documentclass[usegraphicx,usenatbib]{mn2e}
\begin{document}

\title[Cooling of KS~1731$-$260 and MXB~1659$-$29]{Cooling of the quasi-persistent neutron star X-ray transients KS~1731$-$260 and MXB~1659$-$29}

\author[Cackett et~al.]
 {Edward~M.~Cackett$^1$\thanks{emc14@st-andrews.ac.uk},
 Rudy Wijnands$^2$, Manuel Linares$^2$, Jon~M.~Miller$^3$,
 \newauthor Jeroen Homan$^4$, Walter~H.~G. Lewin$^4$
\\ $^1$ School of Physics and Astronomy,
	University of St.~Andrews,
	KY16 9SS, Scotland, UK
\\ $^2$ Astronomical Institute `Anton Pannekoek',
        University of Amsterdam, Kruislaan 403, 1098 SJ,
        Amsterdam, the Netherlands
\\ $^3$ University of Michigan, Department of Astronomy, 500 Church Street, Dennison 814, Ann Arbor, MI 48105
\\ $^4$ MIT Kavli Institute for Astrophysics and Space Research,
        77 Massachusetts Avenue, Cambridge, MA 02139, USA	 
}
\date{Received ; in original form }
\maketitle

\begin{abstract}

We present {\it Chandra} and {\it XMM-Newton} X-ray observations that monitor the neutron star cooling of the quasi-persistent neutron star X-ray transients KS~1731$-$260 and MXB~1659$-$29 for approximately 4 years after these sources returned to quiescence from prolonged outbursts. In both sources the outbursts were long enough to significantly heat the neutron star crust out of thermal equilibrium with the core. We analyse the X-ray spectra by fitting absorbed neutron star atmosphere models to the observations.  The results of our analysis strengthen the preliminary findings of Wijnands et al. that in both sources the neutron star crust cools down very rapidly suggesting it has a high heat conductivity and that the neutron star core requires enhanced core cooling processes.  Importantly, we now detect the flattening of the cooling in both sources as the crust returns to thermal equilibrium with the core. We measure the thermal equilbrium flux and temperature in both sources by fitting a curve that decays exponentially to a constant level.  The cooling curves cannot be fit with just a simple exponential decay without the constant offset.  We find the constant bolometric flux and effective temperature components to be $(9.2 \pm 0.9) \times10^{-14}$ ergs cm$^{-2}$ s$^{-1}$ and $70.0 \pm 1.6$ eV in KS~1731$-$260 and $(1.7 \pm 0.3) \times10^{-14}$ ergs cm$^{-2}$ s$^{-1}$ and $51.6 \pm 1.4$ eV in MXB~1659$-$29.  We note that these values are dependent on the assumed distance to the sources and the column density which was tied between the observations due to the low number of photons in the latter observations.  However, importantly, the shape of the cooling curves is independent of the distance assumed.  In addition, we find that the crust of KS~1731$-$260 cools faster than that of MXB~1659$-$29 by a factor of $\sim$2, likely due to different crustal properties.  This is the first time that the cooling of a neutron star crust into thermal equilibrium with the core has been observed in such detail.

\end{abstract}


\section{Introduction}
Low-mass X-ray binaries consist of either a neutron star or a black hole that accretes matter from a low-mass companion star, where mass transfer occurs because the donor star overflows its Roche-lobe.  The soft X-ray transients form a sub-group of low-mass X-ray binaries.  The vast majority of their time is spent in a quiescent state where little accretion occurs, resulting in X-ray luminosities $< 10^{34}$ ergs s$^{-1}$. However, during outbursts (which usually last weeks to months) the mass-accretion rate rises significantly, with a corresponding large increase in brightness - the typical outburst luminosity is in the range $10^{36} - 10^{39}$ erg s$^{-1}$ \citep[for reviews of neutron star and black hole transients see, e.g.,][]{campanaetal98a,remillardmcclintock06}.

The quiescent X-ray spectra of neutron-star X-ray transients can generally be characterised by two components - a soft, thermal component that dominates below a few keV, and a hard power-law tail that dominates above a few keV.  The power-law component is not always present and its origin is poorly understood, though in some systems it completely dominates \citep[e.g. EXO~1745$-$248 in the globular cluster Terzan~5,][]{wijnandsetal05}.  Several emission mechanisms have been proposed to explain the quiescent X-ray emission, for example, residual accretion onto the neutron star magnetosphere or pulsar shock emission \citep[e.g.,][]{campanaetal98a,menou99,campanastella00,menoumcclintock01}.  However, the mechanism most often used to explain the thermal X-ray emission is that in which the emission is due to the cooling of a hot neutron star which has been heated during outburst \citep{vanparadijsetal87,BBR98}.  In the \citet*{BBR98} model, the neutron star core is heated by nuclear reactions occurring deep in the crust due to accretion during outburst, and this heat is released as thermal emission during quiescence.  Thus, the temperature of the neutron star core and surface in quiescence is then set by the time-averaged mass accretion rate and the neutron star properties.  Therefore, if the time-averaged accretion rate can be estimated, studying quiescent neutron star systems can give important information about the physical processes at work in the very high densities present in neutron star cores.

Some X-ray transients (the quasi-persistent transients) spend an unusually long time in outburst, with outbursts lasting years to decades rather than the more common weeks to months.  For the neutron-star quasi-persistent transients, during outbursts, the crust of the neutron star is heated to a point beyond thermal equilibrium with the stellar interior.  Once accretion falls to quiescent levels, the crust cools thermally, emitting X-rays, until it reaches equilibrium again with the core \citep[e.g.,][]{rutledge02_ks1731}.  Crust cooling curves allow us to probe the properties of neutron stars, such as, the kind of core cooling processes at work (which depends on the equation of state of ultra dense matter) and the heat conductivity of the crust (which likely depends on the properties of the iron lattice of the crust, such as its purity).  While it is expected that the crust can cool significantly between outbursts, the neutron star core should only cool significantly on much longer timescales.  X-ray observations of the two quasi-persistent neutron star X-ray transients KS~1731$-$260 \citep{wijnands2001,wijnands02,wijnandsetal02,rutledge02_ks1731,wijnands05} and MXB 1659$-$29 \citep{wijnandsetal03,wijnandsetal04,wijnands04,wijnands05} have tracked the initial cooling of these neutron stars once they went into a quiescent state.  In this paper, we present additional \textit{Chandra} and \textit{XMM-Newton} observations that further track the cooling of these two neutron stars as they return to thermal equilibrium with the core.

\subsection{KS~1731$-$260}
\begin{figure}
  \centering
  \includegraphics[angle=270,width=8.2cm]{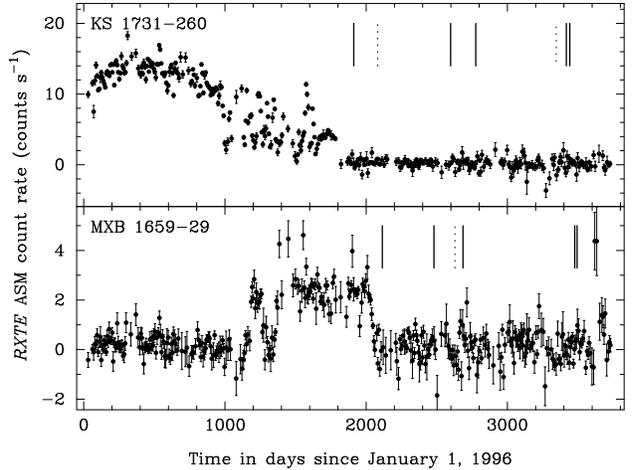}
  \caption{{\it RXTE} All-Sky Monitor 7-day averaged lightcurve for KS~1731$-$260 (top) and MXB~1659$-$29 (bottom).  The times of the {\it Chandra} and {\it XMM-Newton} observations of these sources during quiescence are marked by solid and dotted vertical lines respectively.  The last detection of KS~1731$-$260 in outburst was 1847 days after 1996 January 1, and the last detection of MXB~1659$-$29 was 2076 days after 1996 January 1.}
  \label{fig:rxte_asm_lc}
\end{figure}

KS~1731$-$260 was first detected in outburst in August 1989 by the {\it Mir-Kvant} instrument and further analysis revealed that the source was also found to be active in October 1988 \citep{sunyaev89,sunyaev90a}.  As no previous telescopes had detected this source, it was classified as a soft X-ray transient and the detection of type I X-ray bursts identified the compact object as a neutron star.  Since then, until early 2001, the source remained bright and was observed by many X-ray telescopes including the ART-P and SIGMA telescopes aboard {\it GRANAT}, {\it Ginga, ROSAT, RXTE, ASCA}, and {\it BeppoSAX} \citep{sunyaev90b,barret92,yamauchi90,barret98,smith97,wijnands97,muno00,narita01,kuulkers02}.  In early 2001, KS~1731$-$260 was undetected by monitoring observations by the All-Sky Monitor (ASM) on-board the {\it Rossi X-ray Timing Explorer} ({\it RXTE}) as can be seen by the ASM lightcurve of this source in Fig. \ref{fig:rxte_asm_lc} (top).  Further pointed {\it RXTE} observations with the Proportional Counter Array (PCA) and the bulge scan observations of the Galactic centre region did not detect the source, with the first non-detection by the PCA observations on 2001 February 7 \citep{wijnands2001}.  Using archival {\it RXTE}/PCA data, we find that the final detection of the source with {\it RXTE} was on 2001 January 21, thus it was still in outburst and actively accreting on this date.  We extracted the {\it RXTE} X-ray spectrum of the source on that day and find that a 2-10 keV unabsorbed flux of $\sim2.3 \times 10^{-10}$ ergs cm$^{-2}$ s$^{-1}$ which corresponds to a luminosity of $\sim 1 \times 10^{36}$ erg s$^{-1}$ for a distance of 7 kpc \citep[we assume a distance to KS~1731$-$260 of 7 kpc throughout the paper,][]{muno00}.  Thus, KS~1731$-$260 went into a quiescent state sometime after 2001 January 21 and it was actively accreting for around 12.5 years.  This last detection in outburst gives a much better constraint than had previously been found for when this source turned off.  Unfortunately, there are no other known outbursts of this source, thus we are unable to tell whether this is a typical outburst for this source, or whether previous outbursts have been significantly longer or shorter.

Within two months of KS~1731--260 going into a quiescent phase there was a $\sim$20 ks {\it Chandra} Advanced CCD Imaging Spectrometer (using ACIS-S mode) observation of this source on 2001 March 27 \citep{wijnands2001}.  These authors detected the source at an unabsorbed flux of $\sim2 \times 10^{-13}$ ergs cm$^{-2}$ s$^{-1}$ (0.5-10 keV), which corresponds to a 0.5-10 keV luminosity of $\sim 1 \times 10^{33}$ ergs s$^{-1}$, a factor of $10^3$ lower than at the last {\it RXTE}/PCA detection.  Observations by {\it BeppoSAX} confirmed the low flux of the source \citep{burderi02} reported by \citet{wijnands2001}.  \citet{wijnands2001} note that this luminosity is much lower than expected if this is due to the cooling of the neutron star unless the source spends an unusually long amount of time in a quiescent state (at least several hundred years) and even longer if the crust is dominating emission from the neutron star, or that alternatively, enhanced cooling of the core occurs due to higher levels of neutrino emission.  \citet{rutledge02_ks1731} note that given the long outburst, the crust would be dominating the thermal emission from the neutron star as it would have been heated significantly out of equilibrium with the core.  Given the accretion history of this source \citet{rutledge02_ks1731} simulate the quiescent lightcurve for the source, finding that depending on the crust and core microphysics, the cooling timescale for the crust to return to equilibrium with the core ranges from 1 to 30 years.
 
Additional X-ray observations of this source in quiescence with {\it XMM-Newton} in 2001 September 13-14 revealed that the source had dimmed further to a 0.5-10 keV luminosity of approximately $(2-5) \times 10^{32}$ ergs s$^{-1}$ \citep{wijnandsetal02}. Under the neutron star cooling model, this decrease in luminosity is interpreted as rapid cooling of the the neutron star crust (a factor a $\sim$2-3 in just half a year) suggesting that the crust has a high thermal conductivity and that enhanced core cooling processes are required for the core to be so cool.  Preliminary analysis of two extra {\it Chandra} observations of this source on 2003 February 11 and 2003 August 8 yet again showed a further decrease in the luminosity of the source with a possible flattening of the luminosity cooling curve suggesting that the crust may have returned to equilibrium with the core \citep{wijnands05}.  In this paper we present the final analysis of these {\it Chandra} data, re-analyse all the previous {\it Chandra} and {\it XMM-Newton} observations of this source in quiescence and also present 2 additional {\it Chandra} observations and 1 additional {\it XMM-Newton} observation of this source.

\subsection{MXB~1659$-$29}

MXB~1659$-$29 is an X-ray transient discovered by \citet{lewin76} with {\it SAS-3} in 1976 October.  Type-I X-ray bursts detected by these initial observations established that the compact object was a neutron star.  Between 1976 October and 1978 September, this source was detected several times by both {\it SAS-3} and {\it HEAO 1} \citep{lewin78,share78,griffiths78,doxsey79,cominsky83,cominsky84,cominsky89}.  The source exhibits X-ray eclipses and dips with a period of $\sim7.1$ hr, due to the orbital period \citep{cominsky84,cominsky89}.  In 1979 July, the {\it Hakucho} satellite failed to detect the source \citep{cominsky83}, and later observations by {\it EXOSAT} and {\it ROSAT} \citep{oosterbroek01,verbunt01} failed to detect the source also.  The 0.5-10 keV upper limit on the unabsorbed flux from the {\it ROSAT} observations in 1991 and 1992 is $(1-2) \times 10^{-14}$ ergs cm$^{-2}$ s$^{-1}$ \citep{verbunt01,wijnands02,oosterbroek01}.

The source remained in a quiescent state for a period of 21 years, until on 1999 April 2 {\it BeppoSAX} detected the source in outburst again \citep{intzand_mxb1659_99}.  This outburst was also detected by the {\it RXTE} ASM (see the ASM lightcurve in Fig. \ref{fig:rxte_asm_lc}) and studied in detail by pointed observations with the {\it RXTE} PCA, {\it BeppoSAX} and {\it XMM-Newton} \citep{wachter00,oosterbroek01,sidolietal01,wijnandsetal01,wijnandsetal_mxb1659_02}.

After an outburst of approximately 2.5 years in length, MXB~1659$-$29 went back into a quiescent state in 2001 September.  The last detection of the source was on 2001 September 7 with the {\it RXTE} PCA, and the source was not detected on 2001 September 14, 24, 30 \citep{wijnandsetal_mxb1659_02}.  Just over a month after the source returned to a quiescent state, {\it Chandra} observations, on 2001 October 15, detected it with an unabsorbed 0.5-10 keV flux of $(2.7 - 3.6) \times 10^{-13}$ ergs cm$^{-2}$ s$^{-1}$ which corresponds to a 0.5-10 keV luminosity of $(3.2 - 4.3) \times 10^{33}$ ergs s$^{-1}$ for a distance of 10 kpc \citep{wijnandsetal03}.  This flux is a factor greater than 10 higher than the {\it ROSAT} non-detection upper limit during quiescence in the early 1990s, suggesting that the emission is dominated by thermal emission from the neutron star crust that has been heated significantly out of thermal equilibrium with the core.  However, as in the case of KS~1731$-$260, the flux was lower than expected given the accretion history, suggesting that enhanced cooling processes are occurring in the core.  Further {\it Chandra} observations of MXB~1659$-$29 in quiescence on 2002 October 15 and 2003 May 9 showed that the flux was decaying exponentially, with the source flux decreasing by a factor of 7-9 over the 1.5 years between the 2001 and 2003 {\it Chandra} observations \citep{wijnandsetal04,wijnands04,wijnands05}.  This is consistent with the accretion-heated crust cooling down in quiescence.  The rapid timescale of the cooling implies that the crust may have a large thermal conductivity \citep{wijnandsetal04,wijnands04,wijnands05}.  In this paper we re-analyse all the previous observations of this source in quiescence with {\it Chandra} and also present 2 additional {\it Chandra} observations and 1 additional {\it XMM-Newton} observation of this source.

\section{Observations and Analysis}

\subsection{KS~1731$-$260}

Here we analyse  5 {\it Chandra} and 3 {\it XMM-Newton} observations of KS~1731$-$260 in a quiescent state spread over $\sim$4 years since the end of the outburst in 2001 February.  Details of all the observations are given in Table \ref{tab:ks1731_obs}.   We will first describe the analysis of the {\it Chandra} and then the {\it XMM-Newton} data.
\begin{table*}
\begin{center}
\caption{Details of the {\it Chandra} (CXO) and {\it XMM-Newton} (XMM) observations of KS~1731$-$260.  The {\it XMM-Newton} observations 0137950201 and 0137950301 were so close in time that they were merged together.  Here we give the details for the merged data.  The background-subtracted net count rate is for the 0.5-10 keV band.}
\label{tab:ks1731_obs}
  \begin{tabular}{cccc}
    \hline
     ObsID & Date & Good Time & Net Count Rate\\
     (Telescope) & & (ks) & (counts s$^{-1}$) \\

    \hline
    
    2428 (CXO) & 2001 March 27& 19.4 & $(9.0 \pm 0.7) \times 10^{-3}$ \\
    0137950201/301 (XMM) & 2001 September 13/14& 46.0 (MOS1)& $(1.7 \pm 0.2) \times 10^{-3}$ (MOS1)\\
     & & 46.1 (MOS2) & $(2.0 \pm 0.2) \times 10^{-3}$ (MOS2)\\
     & & 33.8 (pn)   & $(3.7 \pm 0.4) \times 10^{-3}$ (pn)  \\
    3796 (CXO) & 2003 February 11& 29.7 & $(1.4 \pm 0.2) \times 10^{-3}$\\
    3797 (CXO) & 2003 August 8& 29.7 & $(1.1 \pm 0.2) \times 10^{-3}$\\
    0202680101 (XMM) & 2005 February 28/March 1& 70.0 (MOS1) & $(2.4 \pm 1.6) \times 10^{-4}$ (MOS1)\\
     & & 70.0 (MOS2) & $(6.4 \pm 1.7) \times 10^{-4}$ (MOS2) \\
     & & 46.0 (pn)   & $(1.1 \pm 0.4) \times 10^{-3}$ (pn)  \\
    6279 (CXO) & 2005 May 10 & 10.4& $(0.5 \pm 0.3) \times 10^{-3}$\\
    5468 (CXO) & 2005 June 4& 38.5& $(0.8 \pm 0.2) \times 10^{-3}$\\
    \hline
  \end{tabular}
\end{center}
\end{table*}

\subsubsection{\textit{Chandra} analysis}
All the {\it Chandra} observations were taken in ACIS-S configuration.
The {\it Chandra} data were reprocessed and analysed using \textsc{ciao} (v3.3) and \textsc{Caldb} (v3.2.1) following the standard analysis threads.  For all of the {\it Chandra} observations, the source lightcurve and spectrum was extracted from a circle of radius 3\arcsec~around the source position, and the background lightcurve and spectrum was extracted from a source-free annulus with inner radius 7\arcsec~and outer radius 25\arcsec.  We checked the background lightcurve for significant background flares, and none were found.  The background-subtracted net count rates in the 0.5-10 keV band are given in Table \ref{tab:ks1731_obs}.

\subsubsection{\textit{XMM-Newton} analysis}

The first two {\it XMM-Newton} observations were separated by only a matter of hours, therefore we combine both of these together to increase sensitivity \citep[as did][]{wijnandsetal02}.  Here we analyse data from the three EPIC instruments on-board {\it XMM-Newton} - the two metal oxide semiconductor (MOS) cameras and the pn camera which were operated in full-frame mode with the thin filter in all the observations.  To analyse the {\it XMM-Newton} data we use the Science Analysis Software (\textsc{sas}, version 6.5.0).  We extract source lightcurves and spectra from a circle of radius 10\arcsec.  The background extraction region was chosen to be a source-free circle of radius 1\arcmin~close to the source position.  An annulus was not used because of the location of a nearby source, 2MASSI~J173412.7-260548, approximately 30\arcsec away.  Although this source is also present in the {\it Chandra} data the significantly better angular resolution of {\it Chandra} allows the use of an annulus for background extraction. 

In the merged observations (ObsID 0137950201/301) background flares were present.  We kept all data where the count rate for PI $>$ 10 keV was less than 10 counts s$^{-1}$ for both the MOS and pn.  In the last {\it XMM-Newton} observation (ObsID 0202680101) background flares were also present.  In this case we kept all data where the count rate for PI $>$ 10 keV was less than 10 counts s$^{-1}$ for the MOS and less than 15 counts s$^{-1}$ for the pn.  We note that in this last observation the source is not obviously present when viewing an image in the full energy range.  It only becomes apparent when restricting the energy range to between 1 and 2 keV where the vast majority of the photons from the source are emitted during this observation.  The net count rates for these observations are given in Table \ref{tab:ks1731_obs}.

\subsubsection{Spectral analysis}

For spectral analysis of the data we use {\sc xspec} \citep{arnaud96}.  The data was fit with an absorbed neutron star hydrogen atmosphere model \citep*[nonmagnetic case,][]{zavlinetal96,pavlov91} with the mass and radius fixed at the canonical values (1.4 M$_\odot$ and 10 km respectively).  Both the {\it Chandra} and {\it XMM-Newton} data were fit simultaneously with the column density tied between the observations and the temperature left as a free parameter.  The column density was tied between the observations as there were too few counts in the last observations to constrain this parameter.  The parameters for the EPIC MOS1, MOS2 and pn spectra for each {\it XMM-Newton} observation were all tied.   The normalisation of the neutron star atmosphere model was fixed for a distance of 7 kpc \citep[distance determined by][]{muno00}.  Although this distance is uncertain, if this parameter is left free it goes to an unrealistic value that would put the distance to the source at $\sim 20$ kpc (although the error on the distance is very large and includes the assumed 7 kpc distance).  Hence, we chose to fix the distance at 7 kpc.

The data were unbinned, due to the low number of counts in the last few observations, and the W statistic \citep{wachter79} was used in the spectral fitting process with background subtraction.  The W statistic is adapted from the Cash statistic \citep{cash79}, but allows background subtraction. We note that spectral fitting to the first {\it Chandra} observation using both $\chi^2$ and W-statistics resulted in the same parameters to within the errors.  To determine the bolometric flux we extrapolated the model to the energy range 0.01 - 100 keV, which gives approximate bolometric fluxes.  The results of the spectral fits are detailed in Table \ref{tab:ks1731}. The spectral parameters for the first {\it Chandra} and {\it XMM-Newton} observations are consistent with those determined by \citet{wijnandsetal02}.    We also note that we get results consistent with the previously published data for KS~1731$-$260 \citet{wijnandsetal02} when using alternative models to the NSA model (i.e., a blackbody model).
\begin{table*}
\begin{center}
\caption{Neutron star atmosphere model fits to the X-ray spectrum of KS~1731$-$260 for 5 {\it Chandra} (CXO) and 3 {\it XMM-Newton} (XMM) observations.  Luminosity is calculated assuming a distance to the source of 7 kpc. 1-$\sigma$ errors on the parameters are given.  The Modified Julian Date (MJD) given correspond to the mid-point of the observation.  Note the column density parameter is tied between the observations when simultaneously fitting the data, and we determine $N_H = (1.3 \pm 0.1) \times 10^{22}$ cm$^{-2}$.}
\label{tab:ks1731}
  \begin{tabular}{ccccc}
    \hline
     ObsID & MJD & kT$_{\mathrm{eff}}^{\infty}$ & Bolometric flux & Luminosity\\
    (Telescope) &  & (eV) & ($10^{-13}$ ergs cm$^{-2}$ s$^{-1}$) & (10$^{32}$ ergs s$^{-1}$)\\
    \hline
    
    2428 (CXO) & 51995.1  & $103^{+2}_{-3}$ & $4.2^{+0.3}_{-0.5}$ & $24^{+2}_{-3} $ \\
    0137950201/301 (XMM) & 52165.7 & $88 \pm 2$ & $2.2 \pm 0.2$ & $13 \pm 1$\\
    3796 (CXO) & 52681.6  & $76 \pm 3$ & $1.2 \pm 0.2$ & $7 \pm 1$ \\
    3797 (CXO) & 52859.5  & $72 \pm 4$ & $1.0 \pm 0.2 $ & $6 \pm 1$ \\
    0202680101 (XMM) & 53430.5 & $70 \pm 3$ & $0.9^{+0.2}_{-0.1}$ & $5 \pm 1$  \\
    6279 (CXO) & 53500.4  & $67^{+6}_{-9}$ & $0.8 \pm 0.3$ & $4 \pm 2$  \\
    5468 (CXO) & 53525.4  & $70 \pm 3$ & $0.9 \pm 0.2$ &  $5 \pm 1$\\
    \hline
  \end{tabular}
\end{center}
\end{table*}

\subsubsection{Cooling curves}
\begin{figure}
  \centering
  \includegraphics[angle=270,width=7.5cm]{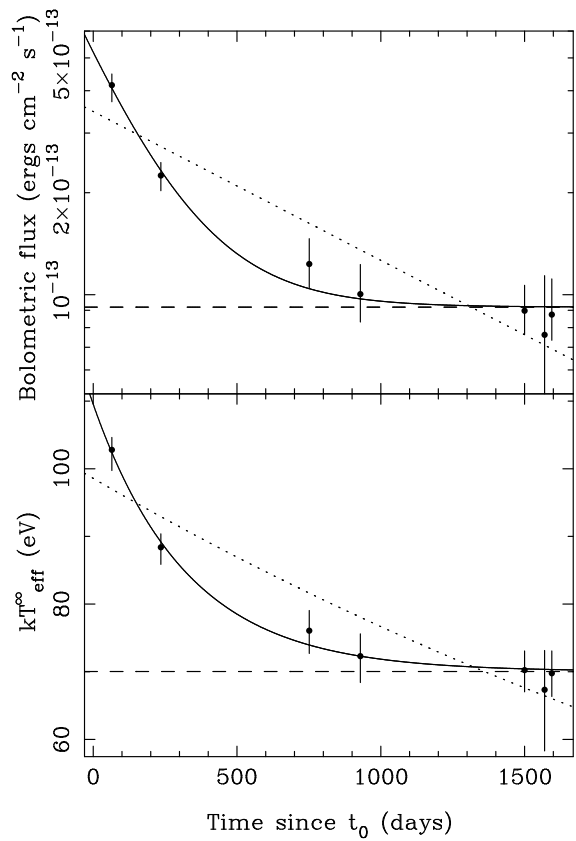}
  \caption{Cooling curves for KS1731$-$260.  {\it Top}:  Bolometric flux versus time since $t_0$, the last detection of the source accreting.  The solid line shows the best fitting exponential decay to a constant.  The constant is indicated as a dashed line.  The dotted line shows the best fitting simple exponential decay, which does not fit the data well.  {\it Bottom}:  Effective temperature for an observer at infinity versus time since the end of the outburst.  The lines are as above.}
  \label{fig:ks1731_curves}
\end{figure}

It is clear from Table \ref{tab:ks1731} that both the effective temperature for an observer at infinity ($T_{\mathrm{eff}}^{\infty}$) and the bolometric flux $F_{\mathrm{bol}}$ decrease significantly with time.   This cooling cannot be fit by a simple exponential decay, giving reduced $\chi^2$ values of 4.3 and 4.8 for fits to the temperature and flux curves (see the dotted curves in Fig. \ref{fig:ks1731_curves}.  However, it is fit well by an exponential decay that levels off to a constant offset of the form $y(t) = a\exp{[-(t-t_0)/b]} + c$, with $a$ a normalisation constant, $b$ the $e$-folding time, $c$ a constant offset, and $t_0$ the start time.  When fitting to the data, $t_0$ was fixed to midday on the last day that the source was observed to be active, MJD 51930.5, though we find that the other parameters are not very sensitive to the exact value of $t_0$.  The best-fitting cooling curves are shown in Fig. \ref{fig:ks1731_curves}.  For the $T_{\mathrm{eff}}^{\infty}$ curve $a = 39.5 \pm 3.6$ eV, $b = 325 \pm 101$ days, and $c = 70.0 \pm 1.6$ eV, with $\chi^2_\nu = 0.2$.  For the $F_{\mathrm{bol}}$ curve $a = (4.3 \pm 0.5) \times10^{-13}$ ergs cm$^{-2}$ s$^{-1}$, $b = 212 \pm 60$ days, and $c = (9.2 \pm 0.9) \times10^{-14}$ ergs cm$^{-2}$ s$^{-1}$, with $\chi^2_\nu = 0.4$.  The reason for the difference between the $e$-folding times of the flux and effective temperature curves is explained in Appendix \ref{ap:A}.

\subsection{MXB~1659$-$29}

We analyse 5 {\it Chandra} observations and 1 {\it XMM-Newton} observation of MXB~1659$-$29 whilst the source was in a quiescent state spanning a period of $\sim$4 years after the end of the outburst in September 2001.  Details of the observations are given in Table \ref{tab:mxb1659_obs}.   We will first describe the {\it Chandra} and then the {\it XMM-Newton} data reduction and analysis. 
\begin{table*}
\begin{center}
\caption{Details of the {\it Chandra} (CXO) and {\it XMM-Newton} (XMM) observations of MXB~1659$-$29. The background-subtracted net count rate is for the 0.5-10 keV band.}
\label{tab:mxb1659_obs}
  \begin{tabular}{cccc}
    \hline
     ObsID & Date & Good Time & Net Count Rate\\
     (Telescope) & & (ks) & (counts s$^{-1}$) \\

    \hline
    
    2688 (CXO) & 2001 October 15 & 17.9 & $(5.2 \pm 0.2) \times 10^{-2}$ \\
    3794 (CXO) & 2002 October 15 & 26.1 & $(9.8 \pm 0.6) \times 10^{-3}$\\
    0153190101 (XMM) & 2003 March 13 & 65.3 (MOS1) & $(3.2 \pm 0.3) \times 10^{-3}$ (MOS1)\\
     & & 65.4 (MOS2) & $(2.9 \pm 0.3) \times 10^{-3}$ (MOS2) \\
     & & 47.1 (pn)   & $(1.1 \pm 0.1) \times 10^{-2}$ (pn)  \\
    3795 (CXO) & 2003 May 9 & 26.2 & $(3.9 \pm 0.4) \times 10^{-3}$\\
    5469 (CXO) & 2005 July 8 & 27.7& $(1.1 \pm 0.2) \times 10^{-3}$\\
    6337 (CXO) & 2005 July 25 & 17.8& $(0.7 \pm 0.2) \times 10^{-3}$\\
    \hline
  \end{tabular}
\end{center}
\end{table*}

\subsubsection{Chandra analysis}
All the {\it Chandra} observations of this source were taken in the ACIS-S configuration.  As in the analysis of the {\it Chandra} data for KS~1731$-$260, we reprocess and analyse the data using \textsc{ciao} (v3.3), \textsc{Caldb} (v3.2.1) and the standard analysis threads.  For all of the {\it Chandra} observations, the source lightcurve and spectrum was extracted from a circle of radius 3\arcsec~around the source position, and the background lightcurve and spectrum was extracted from a source-free annulus with inner radius 7\arcsec~and outer radius 22\arcsec.  We checked the background lightcurve for significant background flares, and none were found.

The analysis of the data for MXB~1659$-$29 is complicated by the fact that this source is eclipsing with an eclipse duration of $\sim$900~s and period of 7.1~hr \citep{cominsky84,cominsky89,wachter00,oosterbroek01}, and so we receive no (or minimal at most) counts from the source during the eclipse in quiescence \citep{wijnandsetal03}.  While there are enough counts in the first {\it Chandra} observation to detect the eclipse in the lightcurve \citep{wijnandsetal03}, this is impossible with later observations and so we manually reduce the exposure time by 900~s to compensate for the source being in eclipse, having checked that only one eclipse occurs during each observation using the ephemeris of \citet{oosterbroek01}.  The background-subtracted net count rates in the 0.5-10 keV band are given in Table \ref{tab:mxb1659_obs}.

\subsubsection{XMM-Newton analysis}
To analyse the {\it XMM-Newton} observation of MXB~1659$-$29 we use {\sc sas} (version 6.5.0).  We analyse data from the EPIC MOS1, MOS2 and pn cameras which were operated in full-frame mode with the medium filter (the source is too faint to be detected on the RGS).  The source lightcurve and spectrum was extracted from a circle with a 15\arcsec radius.  The background lightcurve and spectrum was extracted from a source-free annulus with inner radius of 40\arcsec and outer radius of 100\arcsec.  Several background flares were present during the observation.  We chose to keep all data where the count rate for PI $>$ 10 keV was less than 7 counts s$^{-1}$ for the MOS and less than 10 counts s$^{-1}$ for the pn detector.  Using the ephemeris of \citet{oosterbroek01} we determined that 3 eclipses occurred during the good time of the MOS observations, and so we manually reduce the exposure times for the MOS observations by 2.7 ks.  In the pn observations several flaring events occurred during the eclipse times.  We determine the total amount of good time for the pn during the eclipses to be 1.675 ks, and thus we reduce the exposure time by this amount.  The net count rates for this {\it XMM-Newton} observation are given in Table \ref{tab:mxb1659_obs}.

\subsubsection{Spectral analysis}
We perform a similar spectral analysis for the MXB~1659$-$29 data as for the KS~1731$-$260.  The {\it Chandra} and {\it XMM-Newton} spectra are left unbinned and fit simultaneously with an absorbed non-magnetic neutron star hydrogen atmosphere model using the W statistic with background subtraction. The column density is tied between the separate observations, the mass and radius fixed at their canonical values, and the temperature was the only parameter left free between the observations.  The parameters are tied between the MOS1, MOS2 and pn spectra of the {\it XMM-Newton} observation.  The distance to this source is in the range 10-13 kpc \citep{muno01,oosterbroek01}.  When leaving the normalisation as a free parameter in the {\it Chandra} fits this leads to a distance to the source of 7.8 kpc with no constraint on the upper limit of the distance and a lower limit of 6.2 kpc.  So that the uncertainty in the distance does not dominate the uncertainties in the other parameters, we fix the normalisation of the neutron star atmosphere for a distance of 10 kpc.  Spectral parameters determined by these fits are detailed in Table \ref{tab:mxb1659}.  The parameters we find for the first three {\it Chandra} observations which were previously analysed by \citet{wijnandsetal04} are consistent with the values these authors determined when assuming the same distance.
\begin{table*}
\begin{center}
\caption{Neutron star atmosphere model fits to the X-ray spectrum of MXB~1659$-$29 for 5 {\it Chandra} (CXO) and 1 {\it XMM-Newton} (XMM) observations.  Luminosity is calculated assuming a distance to the source of 10 kpc.  1-$\sigma$ errors on the parameters are given.  The MJD given correspond to the mid-point of the observation.   Note the column density parameter is tied between the observations when simultaneously fitting the data, and we determine $N_H = (0.20 \pm 0.02) \times 10^{22}$ cm$^{-2}$.}
\label{tab:mxb1659}
  \begin{tabular}{ccccc}
    \hline
     ObsID & MJD & kT$_{\mathrm{eff}}^{\infty}$ & Bolometric flux & Luminosity\\
    (Telescope) &  & (eV) & ($10^{-14}$ ergs cm$^{-2}$ s$^{-1}$) & (10$^{32}$ ergs s$^{-1}$)\\
    \hline
    
2688 (CXO) & 52197.8  & $121 \pm 2$ & $41\pm3$& $49\pm3$\\
3794 (CXO) & 52563.2  & $85 \pm 2$ & $10 \pm 1$ & $12\pm1$\\
0153190101 (XMM) & 52712.2 & $77 \pm 1$ & $6.6^{+0.4}_{-0.5}$ & $7.9^{+0.5}_{-0.6}$ \\
3795 (CXO) & 52768.9 & $73^{+2}_{-1}$ & $5.1^{+0.6}_{-0.4}$ & $6.1^{+0.7}_{-0.4}$ \\
5469 (CXO) & 53560.0 & $58^{+3}_{-4}$ & $2.0^{+0.4}_{-0.5}$ & $2.4^{+0.5}_{-0.6}$ \\
6337 (CXO) & 53576.7 & $54^{+4}_{-5}$ & $1.5\pm0.5$ & $1.8\pm0.6$ \\
    \hline
  \end{tabular}
\end{center}
\end{table*}

\subsubsection{Cooling curves}
\begin{figure}
  \centering
  \includegraphics[angle=270,width=7.6cm]{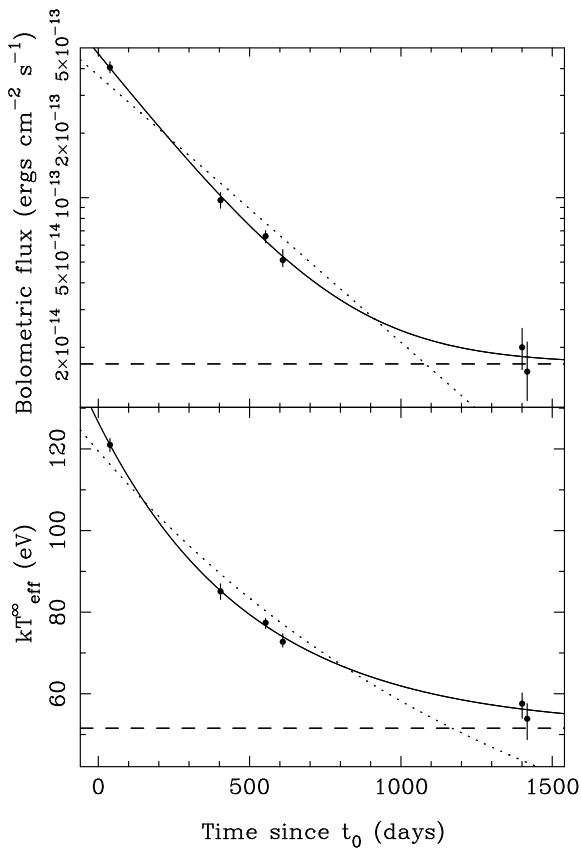}
  \caption{Cooling curves for MXB1659$-$29. {\it Top}:  Bolometric flux versus time since $t_0$, the last detection of the source accreting.  The solid line shows the best fitting exponential decay to a constant.  The constant is indicated as a dashed line. The dotted line shows the best fitting simple exponential decay, which does not fit the data well.  {\it Bottom}:  Effective temperature for an observer at infinity versus time since the end of the outburst.  The lines are as above.}
  \label{fig:mxb1659_curves}
\end{figure}

From the spectral fitting we find that both $T_{\mathrm{eff}}^{\infty}$ and $F_{\mathrm{bol}}$ decrease significantly with time after the end of the outburst.  We measure an unabsorbed 0.5 - 10 keV flux of $6.3^{+2.7}_{-2.5} \times 10^{-15}$ ergs cm$^{-2}$ s$^{-1}$ during the last {\it Chandra} observation which is now lower than the upper limit from the previous non-detection with {\it ROSAT} in the early 1990s when the source had been in quiescence for over 10 years.  As with KS~1731$-$260 this cooling cannot be fit by a simple exponential decay curve, giving reduced $\chi^2$ values of 17.5 and 13.6 for fits to the temperature and flux curves (see dotted line in Fig.\ref{fig:mxb1659_curves}) .  The cooling curves require an exponential decay which levels off to a constant of the form $y(t) = a\exp{[-(t-t_0)/b]} + c$.  When fitting to the data, $t_0$ was fixed to midday on the last day that the source was observed to be active, MJD 52159.5.  The best-fitting cooling curves are shown in Fig. \ref{fig:mxb1659_curves}.  For the $T_{\mathrm{eff}}^{\infty}$ curve $a = 75.0 \pm 2.8$ eV, $b = 505 \pm 59$ days, and $c = 51.6 \pm 1.4$ eV, with $\chi^2_\nu = 0.5$.  For the $F_{\mathrm{bol}}$ curve $a = (4.5 \pm 0.2) \times10^{-13}$ ergs cm$^{-2}$ s$^{-1}$, $b = 242 \pm 13$ days, and $c = (1.7 \pm 0.3) \times10^{-14}$ ergs cm$^{-2}$ s$^{-1}$, with $\chi^2_\nu = 0.5$.

\section{Discussion}

We have presented {\it Chandra} and {\it XMM-Newton} observations monitoring two quasi-persistent neutron star X-ray transients (KS~1731$-$260 and MXB~1659$-$29) in a quiescent state.  The long accretion episodes onto both of these neutron stars before they went into quiescence significantly heated the neutron star crust out of equilibrium with the core.  The monitoring observations cover a period of $\sim$4 years since the sources returned to quiescence.  Spectral fitting with an absorbed neutron star hydrogen atmosphere model to the data has clearly shown both of them cooling exponentially over time.  From the initial results of this monitoring \citep{wijnandsetal02,wijnandsetal04} and the preliminary results of \citet{wijnands05} it was not clear whether either of the neutron star crusts had reached thermal equilibrium with the cores.  However, the results of this latest monitoring indicates that KS~1731$-$260 and MXB~1659$-$29 have now reached equilibrium again and we have been able to measure the base level of the flux and effective temperature of these sources resulting from the state of the hot core.  The additional data in this current work provides a significant improvement on the previous work with respect to constraints on the cooling curve.

KS~1731$-$260 initially decreased by a factor of 2 in bolometric flux over the first half a year.  Over the $\sim$4 years since the source went into quiescence it has decreased in flux by a factor of $\sim$5. It reached this level $\sim$2.5 years after the end of the outburst and has remained at it since then, indicating that the crust returned to equilibrium with the core in this amount of time.  This base level is set by the temperature of the core, which depends on the time-averaged mass accretion rate onto the source over thousands of years.  Fitting an exponential decay with a constant offset to the data we get the equilibrium temperature to be $70.0 \pm 1.6$ eV, and the base bolometric flux level due to thermal emission from the core as $(9.2 \pm 0.9) \times 10^{-14}$ ergs cm$^{-2}$ s$^{-1}$, which corresponds to a luminosity of $(5.4 \pm 0.5) \times 10^{32}$ ergs s$^{-1}$.  The $e$-folding times of the exponential decay were determined to be $325 \pm 101$ days for the temperature cooling curve and $212 \pm 60$ days for the flux cooling curve. 

MXB~1659$-$29 decreased by a factor of 4 in bolometric flux over the first year after going into quiescence, and by a factor of 24 over the $\sim$4 years of monitoring.  \citet{wijnandsetal04} found no evidence after the first 3 {\it Chandra} observations of this source that the temperature or flux was reaching a equilibrium level set by the temperature of the core.  The additional last 2 {\it Chandra} observations presented here indicate that the rate of cooling has decreased.  Fitting an exponential decay with a constant offset to the data reveals that it has reached the base level which we determine to be $51.6 \pm 1.4$ eV for the temperature and $(1.7 \pm 0.3) \times 10^{-14}$ ergs cm$^{-2}$ s$^{-1}$ for the bolometric flux, which corresponds to a luminosity of $(2.0 \pm 0.4) \times 10^{32}$ ergs s$^{-1}$ (assuming d=10 kpc).  The $e$-folding times of the exponential decay were determined to be $505 \pm 59$ days for the temperature cooling curve and $242 \pm 13$ days for the flux cooling curve.

\citet{rutledge02_ks1731} calculated detailed cooling curves for KS~1731$-$260 predicting the behaviour of the quiescent thermal emission from the neutron star given the source's accretion history (see their fig. 3).  Their models compare how the emission could evolve for standard versus enhanced core cooling and low versus high crust thermal conductivity.  These authors find that regardless of the core cooling, if the crust has a low thermal conductivity, then the crust should dominate emission over the core for $\sim$30 years, after which time there is a large drop in the source luminosity.  For high-conductivity crusts the timescale for this transition is much shorter, $\sim$1 year.  Comparing these timescales to the observed cooling of KS~1731$-$260, then, in the context of these models, the crust in this source must have a high thermal conductivity \citep{wijnandsetal02,wijnands05}, suggesting that it may be made of a pure iron lattice as increasing levels of impurity lower the conductivity.  Unfortunately no detailed cooling curves have been calculated for MXB~1659$-$29.  As the cooling is dependent on the accretion history of the source, we cannot directly compare the cooling of MXB~1659$-$29 with the model curves calculated for KS1731$-$260 by \citet{rutledge02_ks1731}.  However, in these models to see a rapid drop in flux within a few years of the source going into quiescence the crust must have a high conductivity, though there may be other physical processes in the crust that allow it to cool rapidly.  In particular, we note that the \citet{rutledge02_ks1731} models do not include Cooper-pair neutrino emission in the crust which will change the behaviour significantly \citep*[e.g.,][]{yakovlev99}.  See the discussion in \citet{jonker06} for further comment on this difference.

Once the crust has cooled, the quiescent luminosity is set by the emission from the hot core \citep{rutledge02_ks1731}.  In the standard core cooling mechanisms the heat lost by neutrino emission is negligible and thus the heat deposited in the star is radiated as thermal emission.  In the case where enhanced core cooling is allowed, the enhanced level of neutrino emission, via the direct Urca process or Cooper-pairing, means that the vast majority of heat deposited escapes as neutrinos \citep{colpietal01,ushomirskyrutledge01,rutledge02_ks1731,gusakov04}.  If enhanced core cooling occurs, then the quiescent luminosity set by the core temperature is significantly lower than if only standard core cooling occurs.  We note that continued cooling is predicted in the \citet{rutledge02_ks1731} models after the crust and core are in equilibrium.  However, this cooling is at a significantly slower rate.
 
Importantly, in both sources we have now measured the thermal equilibrium temperature, which is set by the core temperature, for the first time. The quiescent luminosity at the equilibrium level of both these sources is quite typical of the quiescent emission from other `normal' X-ray transients \citep[e.g. see fig. 5 in][]{jonker04} whose outbursts are not as long as those of either source and last usually for only weeks to months.  For the core of KS~1731$-$260 and MXB~1659$-$29 to have a similar temperature, and hence quiescent luminosity, to the other X-ray transients then either the time-averaged mass accretion rate has to be similar to the other X-ray transients, or enhanced levels of core cooling are required.  For the time-averaged mass accretion rates to be similar to other X-ray transients this would require extremely long periods of quiescence of the order of hundreds to thousands of years if the observed outbursts are typical.  For MXB~1659$-$29 there was only a period of 21 years between outbursts and both known outbursts lasted for several years.  The long-timescale accretion history for KS~1731$-$260 is less clear with only one outburst having been observed, however, extremely long quiescent periods are hard to explain with disk instability models \citep[e.g.][]{lasota2001}.  The measurement of the equilibrium level for both these sources confirms the previous findings by \citep{wijnandsetal02,wijnandsetal04,wijnands05} from initial monitoring that enhanced core cooling processes are required for these sources.

Comparing the $e$-folding timescales of the two sources, we find that KS~1731$-$260 cools at a faster rate than MXB~1659$-$29 (by a factor of $\sim$1.5), suggesting that the heat conductivity of the crust in KS~1731$-$260 is greater than that in MXB~1659$-$29.   This interesting difference is likely due to differences in the properties of the crust between the two sources.  In the \citep{rutledge02_ks1731} models this would mean that KS~1731$-$260 has less impurities in the crust compared to MXB~1659$-$29 as a pure iron crust has a higher conductivity than one containing heavier metals also.  However, one might expect the opposite to be the case, given that KS~1731$-$260 was actively accreting for 5 times longer than MXB~1659$-$29. 

Given that we now have accurate cooling curves for these sources, and know the equilibrium level temperature, which places limits on the long-term time-averaged mass accretion rate, theoretical models need to be fit to this data to determine how high the thermal conductivity needs to be and what levels of neutrino emissions are really needed, or any alternative properties that can account for the rapid cooling of the crust.  As more than one outburst has been observed for MXB~1659$-$29, and because the cooling curve is much better constrained, we suggest that this source is possibly a better object to determine accurate theoretical cooling curves for than KS~1731$-$260. 

Although some neutron-star transients do not require enhanced core cooling to explain their quiescent luminosity e.g. Aql~X$-$1, 4U~1608$-$522 \citep{colpietal01}, XTE~J2123$-$058 \citep[assuming the recurrence timescale is $>$70 years,][]{tomsick04} and SAX~J1748.8$-$2021 in the globular cluster NGC~6440 \citep{cackett2005}, there are several neutron-star transients in addition to KS~1731$-$260 and MXB~1659$-$29 that require either very-low time-averaged mass accretion rates or enhanced core cooling to explain their quiescent luminosity e.g. Cen X-4 \citep{colpietal01}, X1732$-$304 in the globular cluster Terzan~1 \citep{WHG02, cackett06}, RX~J170930.2$-$263927 \citep{jonker03}, SAX~J1810.8$-$2609 \citep{jonkerwijnandsvanderklis04}, EXO~1747$-$214 \citep{tomsick05} and 1H~1905+000 \citep{jonker06}.  While the quiescent luminosity of KS~1731$-$260 and MXB~1659$-$29 is comparable to many quiescent neutron star transients, it is significantly higher than the emerging group of quiescent neutron star transients that are very faint in quiescence, with quiescent luminosities of less than 10$^{32}$ ergs s$^{-1}$.  Although these faint quiescent sources include the millisecond X-ray pulsars SAX~J1808.4$-$3658 \citep{campanaetal02}, XTE~J0929$-$314, XTE~J1751$-$305 \citep{wijnandsetal05b} and XTE~J1807$-$294 \citep{campanaetal05} which may have different properties due to higher magnetic fields, there are also a number of `normal' neutron-star X-ray transients that fall into this category (SAX~J1810.8$-$2609,  XTE~J2123$-$058, EXO~1747$-$214). A notable faint quiescent source is the quasi-persistent neutron-star transient 1H~1905+000 as although it was seen to be in outburst for at least 11 years in the 1970s/80s, a recent {\it Chandra} observation did not detect the source in quiescence, with an upper limit of $1.8 \times 10^{31}$ ergs s$^{-1}$ \citep[0.5-10 keV; for a distance of 10 kpc,][]{jonker06}.  Apart from KS~1731$-$260, MXB~1659$-$29 and 1H~1905+000, the only other quasi-persistent source to have gone into quiescence is X1732$-$304 in the globular cluster Terzan~1.  Using recent {\it Chandra} observations of Terzan~1 \citet{cackett06} find that the most likely quiescent counterpart to this source has a 0.5-10 keV luminosity of $2.6 \times 10^{32}$ ergs s$^{-1}$, about 6 years after the end of an outburst that lasted at least 12 years.  For comparison, the 0.5-10 keV luminosities of the last observations of KS~1731$-$260 and MXB~1659$-$29 are $2.8^{+0.8}_{-0.6} \times 10^{32}$ ergs s$^{-1}$ and $7.5^{+3.2}_{-2.9} \times 10^{31}$ ergs s$^{-1}$ respectively. Therefore, KS~1731$-$260, MXB~1659$-$29 and X1732$-$304 all have higher quiescent luminosities than 1H~1905+000. However, 1H~1905+000 was observed much longer after the end of its outburst then any of the other 3 sources.  While the difference in quiescent luminosity may suggest that these sources will continue to cool further over the next several decades, it is more likely due to differing long-term time-averaged mass accretion rates or neutron star properties.  The differences in the properties of quiescent neutron star transients need to be explored further by additional observations to determine why some sources require enhanced core cooling and why some sources are much fainter than others. Observations of the next quasi-persistent neutron-star transient sources to return to quiescence (EXO~0748$-$676 and GS~1826$-$238 are both quasi-persistent sources that are still active) will help determine whether a decline in flux and temperature to a constant level on the timescale of years, rather than decades, is typical. 

We used an absorbed neutron star atmosphere model to fit to the data.  If we use simple blackbody model in the spectral fitting instead, we get similar cooling curves, yet with different temperatures due to the differences between the shapes of blackbody and neutron star atmosphere spectra.  When performing the spectral fits to both sources the normalisation of the neutron star atmosphere component was fixed to the distance of the sources.  Unfortunately, these distances are quite uncertain, thus, if the distances are wildly different from the values that we have assumed, the temperatures and fluxes that we get from the spectral fits will be incorrect.  However, if the normalisation is left free when fitting, then the uncertainty in this parameter dominates the errors in the other parameters.  Importantly, the timescale of the decrease in the temperature and flux of these sources should be unchanged by a different distance, only the absolute values of the temperature and flux will change.  If the sources are much closer than assumed then the temperature measured will be lower and the column density higher, and the opposite if the sources are further away, as demonstrated by \citet{wijnandsetal04} who fit the first three {\it Chandra} observations of MXB~1659$-$29 assuming 3 different distances to the source ranging from 5 - 13 kpc.

While we have interpreted the observations of KS~1731$-$260 and MXB~1659$-$29 in terms of the crust and core cooling model there are alternative models to explain quiescent neutron star emission including residual accretion or pulsar shock emission \citep[e.g.,][]{campanaetal98a,menou99,campanastella00,menoumcclintock01}.  While these are often used to explain the power-law tail present in the spectra of some neutron star transients, residual accretion can produce soft spectra if accretion occurs onto the neutron star surface \citep{zampierietal95}. However, while variations in the luminosity of the sources in quiescence can be explained by changes in the accretion rate or the amount of matter interacting with the magnetic field, our observations show that this would have to decrease exponentially in a smooth manner over a period of several years.  Given that at the end of the outbursts of `normal' short-duration neutron-star transients they reach their quiescent states within weeks \citep[e.g.][]{campanaetal98b,jonker03} and that variations in accretion rate are generally not smooth, these alternative models seem unlikely.

We note that from the initial observations of both sources it was found that a power-law component (which is often attributed to residual accretion)  in addition to the neutron star atmosphere component was not required to fit the spectra \citep{wijnands2001,wijnandsetal02,wijnandsetal03,wijnandsetal04}.
While it has been observed that the contribution from the power-law component compared to the thermal component increases with decreasing quiescent luminosity below $\sim$10$^{33}$ ergs s$^{-1}$ \citep{jonkerwijnandsvanderklis04,jonker04}, the statistics of the last few observations are unfortunately not good enough to check for this.  However, if residual accretion is occurring during quiescence it should be on-top of the `rock bottom' thermal emission due to deep crustal heating \citep{BBR98,rutledge02_ks1731} that we have now detected.  Several neutron star transients have been observed to vary during quiescence e.g. Aql X$-$1, Cen X$-$4 and SAX J1748.8$-$2021 in the globular cluster NGC~6440 \citep{rutledgeetal01b,rutledgeetal01a,rutledge02,campanastella03,campanaetal04,cackett2005}.  This variability can be interpreted as a variable power-law component, possibly due to changes in the residual accretion rate or the amount of matter interacting with the magnetic field.  Unfortunately this idea is hard to investigate with this data due to the low number of counts in the latter spectra.  However, any variability in the flux from KS~1731$-$260 and MXB~1659$-$29 over the next few years around the base set by the thermal emission from the neutron star may indicate that some residual accretion is occurring.

\subsection*{Acknowledgements}

EMC gratefully acknowledges the support of PPARC.  {\it XMM-Newton} is an ESA science mission with instruments and contributions directly funded by ESA Member States and NASA.  ASM/{\it RXTE} results provided by the ASM/{\it RXTE} teams at MIT and at the RXTE SOF and GOF at NASA's GSFC.

\bibliographystyle{mn2e}
\bibliography{iau_journals,qNS_mnras}

\appendix
\section{Relationship between flux and temperature cooling timescales} \label{ap:A}
Assuming that the X-rays are due to some blackbody-type emission, then the flux, $F$, and temperature, $T$, and just simply related via
\begin{equation}
F = AT^4 \, ,
\label{eq:app1}
\end{equation}
where $A = \frac{R^2}{D^2}\sigma$ and $R$ is the emitting radius, $D$ the distance, and $\sigma$ the Stefan-Boltzmann constant.  We are modelling the cooling of the neutron star surface as an exponential decay to a constant level, i.e.,
\begin{equation}
  T(t) = T_\infty + T^\prime(t)
\label{eq:app2}
\end{equation}
where $T_\infty$ is the temperature at $t = \infty$ which we assume here to be constant, and $T^\prime(t)$ describes the evolution of the temperature of the neutron star crust with time, which we model as:
\begin{equation}
  T^\prime(t) = a\exp{[-(t-t_0)/b]} \, ,
 \label{eq:app3} 
\end{equation}
with $a$ a normalisation constant, $b$ the $e$-folding time, and $t_0$ the start time.  Combining Eq. \ref{eq:app1} and \ref{eq:app2}, leads to
\begin{equation}
\label{eq:app4}
F(t) = AT(t)^4 = A[T_\infty + T^\prime(t)]^4 \,.
\end{equation}
At the start of the cooling, when $T_\infty \ll T^\prime(t)$, then the $T^\prime(t)^4$ term dominates, and Eq. \ref{eq:app4} becomes
\begin{equation}
\label{eq:app5}
F(t) \approx AT^\prime(t)^4 = Aae^{-4(t-t_0)/b} \,.
\end{equation}
and there should be a factor of 4 difference between the $e$-folding times of $F$ and $T$.  This was found to be the case for the initial cooling of MXB~1659$-$29 \citep{wijnandsetal04}.  However, further on in the cooling, once $T^\prime(t) \sim  T_\infty$, then the $e$-folding time of $F$ becomes a complicated function of the $e$-folding time of $T$ due to the constant term.  Finally, once the cooling is negligable and $T^\prime(t) \ll  T_\infty$, then obviously, $F(t) = AT_\infty^4$.

\end{document}